THE EUROPEAN
PHYSICAL JOURNAL C

Regular Article - Theoretical Physics

# Non-singular gravitational collapse through modified Heisenberg algebra

Gabriele Barca[1,2,a] , Giovanni Montani[1,3,b]

[1] Department of Physics, "Sapienza" University of Rome, P.le Aldo Moro, 5, 00185 Rome, Italy
[2] INFN, Sezione di Roma1, P.le A. Moro 2, 00185 Rome, Italy
[3] ENEA, Fusion and Nuclear Safety Department, C.R. Frascati, Via E. Fermi, 45, 00044 Frascati, RM, Italy

**Abstract** We study the effects of cut-off physics, in the form of a modified algebra inspired by Polymer Quantum Mechanics and by the Generalized Uncertainty Principle representation, on the collapse of a spherical dust cloud. We analyze both the Newtonian formulation, originally developed by Hunter, and the general relativistic formulation, that is the Oppenheimer–Snyder model; in both frameworks we find that the collapse is stabilized to an asymptotically static state above the horizon, and the singularity is removed. In the Newtonian case, by requiring the Newtonian approximation to be valid, we find lower bounds of the order of unity (in Planck units) for the deformation parameter of the modified algebra. We then study the behaviour of small perturbations on the non-singular collapsing backgrounds, and find that for certain range of the parameters (the polytropic index for the Newtonian case and the sound velocity in the relativistic setting) the collapse is stable to perturbations of all scales, and the non-singular super-Schwarzschild configurations have physical meaning.

## 1 Introduction

The problem of understanding the final fate of the gravitational collapse of an astrophysical object is a long standing question in literature [1–3]. In particular, the existence of the upper limits for the mass of compact stars [4–7], above which the gravitational collapse is no longer contrasted by the matter pressure with the consequent formation of a black hole, constitutes one of the most outstanding and still debated results [8–10]. Indeed the observation of neutron stars with mass potentially greater than two Solar masses [11–13] opened the way to a series of conjectures concerning the possible physical explanation for this unexpected evidence, including scenarios with new physics for the gravitational field (see for instance the so-called scalarization phenomenon in modified gravity [14–16]).

Here we consider the collapse of a spherical dust cloud, infalling under the effect of its self-gravity, both in the Newtonian and in the fully relativistic limit. The peculiarity of our study is that we introduce features of cut-off physics in the Hamiltonian formulation of the dynamics through modified Poisson brackets, which are the (semi)classical limit of modified commutation relations. This kind of modified algebras, taking the form $[\hat{q}, \hat{p}] = i\hbar F(\hat{p})$, are usually developed to introduce corrections at high energies close to the Planckian regime, and as such are very useful to introduce in a Hamiltonian system corrections from more fundamental Quantum Gravity theories through a simple independent framework. In particular, the approach that we will use here is inspired by the so-called Polymer Quantum Mechanics [17], when expanded in the free cut-off parameter [18,19]. This way, we are including in the gravitational collapse the ingredients for a repulsive-like gravity, similarly to what happens in cosmology when the emergence of a Big Bounce is recovered and singularities are removed (as for example in the frameworks of Loop Quantum Cosmology [20–23], of Group Field Theories [24,25], of Polymer Cosmology [26–31], or of other modified approaches to gravity [32–35]).

In the Newtonian limit, we adopt the representation of the spherical collapse proposed in [36], which consists of a Lagrangian description for the dynamics of the background configuration and of an Euler formulation of the behavior characterizing small perturbations. While the background dynamics is characterized by a pressureless free fall, when studying the perturbations we adopt a polytropic equation of

[a] e-mail: gabriele.barca@uniroma1.it (corresponding author)
[b] e-mail: giovanni.montani@enea.it

Springer

state and the pressure contribution is relevant for the system stability.

In the general relativistic case, we adopt the Oppenheimer-Snyder model [37] in which the region external to the cloud is, according to the Birkhoff theorem [38,39], a Schwarzschild spacetime, while the interior of the collapsing object is associated to a Robertson-Walker geometry with positive curvature. The two spacetime regions are then suitably matched on the boundary of the collapsing cloud. The stability of this collapse is then studied by considering the dynamics of the interior as the background, in agreement with the Lifshitz formulation of the cosmological perturbations [40,41]. The equation of state for these perturbations has been taken in the isothermal form and the constant sound velocity is a free parameter, replacing the polytropic index of the Newtonian formulation.

We stress that the assumption of a free falling background configuration, made both in the non-relativistic and relativistic cases, has been chosen in order to emphasize the effect of the repulsive gravity induced by cut-off physics, simply because they are not hidden here by the presence of a matter pressure contribution.

The present analysis is characterized by two main relevant results. First, it is always possible to obtain an asymptotically static configuration of the background collapse in correspondence to a radius greater than the Schwarzschild value; second, for a suitable range of the free parameters of the perturbation dynamics, the background configuration results to be stable to small perturbations. Furthermore, it is worth stressing that these two outputs of our analysis remain valid in the limit of a very small (even sub-Planckian) value for the cut-off parameter that characterizes the modification of the Poisson brackets. This fact suggests that the presence of a cut-off physics in the gravitational collapse constitutes an intrinsic modification of the gravitational force with respect to the standard Newtonian or Einsteinian gravity and that the singular collapse is never recovered in the modified dynamics.

In other words, the present analysis states that, if we include gravity modifications in the description of a spherical dust collapse, as expected in an effective quantum gravity scenario, the resulting dynamics is always associated to the existence of a physical (super-Schwarzschild) static and, for a given range of the free parameters of the model, stable configuration, i.e. what we could call a stable "dust star".

These results, and in particular the capability of cut-off physics effects to determine a macroscopic modification i.e. the stabilization of the dust collapse above the event horizon, open a new perspective in understanding the basic ingredients to fix the morphology and the final fate of astrophysical bodies. More specifically, once a real equation of state is considered and the star radial inhomogeneity properly accounted for, it could be possible to give constraints on the value of the cut-off parameter that could accommodate the observed violation of the Chandrasekhar or Tolman–Oppenheimer–Volkoff limits.

The paper is organized as follows. In Sect. 2 we introduce the modified algebra as a deformation of the canonical commutation relations, that in the semiclassical limit becomes a deformation of the Poisson brackets. In Sect. 3 we present the Hamiltonian formulation for the classical and modified Hunter model, i.e. the Newtonian model for dust collapse, and in Sect. 4 we introduce perturbations on this background. In Sect. 5 we present the Oppenheimer–Snyder model in its Hamiltonian formulation, and the modified dynamics obtained with the deformed algebra, while in Sect. 6 we introduce perturbations in this relativistic framework. Section 7 concludes the paper with a summary and some remarks.

## 2 Modified Heisenberg algebra

In this section we introduce the modified Heisenberg algebra that we use to implement critical points on the classical evolution, thus solving the gravitational singularities. It is inspired by quantum gravity and quantum cosmological theories such as Polymer Quantum Mechanics (PQM) [17,42] and the Generalised Uncertainty Principle (GUP) representation [43–47].

The algebra takes the form

$$[\hat{q}, \hat{p}] = i\left(1 - \frac{\mu^2 \ell_P^2 \hat{p}^2}{\hbar^2}\right), \qquad (1)$$

where $\hat{q}$ and $\hat{p}$ are two generic conjugate operators and $\mu$ is a real positive deformation parameter descending from the lattice spacing of PQM. In this commutator the necessary fundamental constants appear in order to have the deformation parameter $\mu$ dimensionless, as is sometimes done in GUP literature [48,49]; for example, in this case we assumed that $q$ and $p$ are the standard position and momentum.

Due to the modified commutator depending on $p$, this kind of algebras is usually studied in the momentum polarization, i.e. a representation where wavefunctions $\Psi = \Psi(p)$ are functions of the momentum and the corresponding operator acts multiplicatively on them as $\hat{p}\Psi(p) = f(p)\Psi(p)$. Through a simple procedure introduced in [50], by asking that the operator $\hat{q}$ acts simply differentially as in Standard Quantum Mechanics (SQM), we can find the modified action of the momentum operator as

$$\frac{df}{dp} = 1 - \frac{\mu^2 \ell_P^2 f^2}{\hbar^2}, \quad \frac{\hbar}{\ell_P} \frac{\text{arctanh}(\frac{\mu \ell_P f}{\hbar})}{\mu} = p; \qquad (2)$$

therefore the action of the two fundamental operators is

$$\hat{p}\,\Psi(p) = \frac{\hbar}{\ell_P}\,\frac{\tanh(\frac{\mu\ell_P p}{\hbar})}{\mu}\,\Psi(p), \tag{3a}$$

$$\hat{q}\,\Psi(p) = i\,\hbar\,\frac{\mathrm{d}}{\mathrm{d}p}\Psi(p). \tag{3b}$$

It is trivial to verify that in the limit $\mu \to 0$ these revert to the operators of SQM in the standard momentum polarization; the corrections that this algebra introduces are usually relevant at high energies, i.e. when the $p^2$ term approaches unity.

It is possible to implement these corrections also on a (semi)classical level through an effective theory; in this case, the modified algebra (1) becomes a rule for Poisson brackets:

$$\{q, p\} = 1 - \frac{\mu^2 \ell_P^2 p^2}{\hbar^2}; \tag{4}$$

$$\dot{q} = \{q, \mathcal{H}\} = \frac{\partial \mathcal{H}}{\partial p}\left(1 - \frac{\mu^2 \ell_P^2 p^2}{\hbar^2}\right), \tag{5a}$$

$$\dot{p} = \{p, \mathcal{H}\} = -\frac{\partial \mathcal{H}}{\partial q}\left(1 - \frac{\mu^2 \ell_P^2 p^2}{\hbar^2}\right), \tag{5b}$$

where $\mathcal{H}(q, p)$ is a Hamiltonian function. As mentioned earlier, this kind of equations of motion usually have an additional critical point at $p = \hbar/\mu\ell_P$ and are therefore used to avoid and remove singularities in cosmological models [19,51].

We will use this semiclassical formulation to study the collapse of a dust cloud, both in a Newtonian and in a Relativistic setting.

## 3 Newtonian gravitational collapse

In this section we introduce the Newtonian description for the collapse of a dust cloud, first developed by Hunter [36], in its Hamiltonian formulation. The Hunter model consists in a homogeneous and isotropic sphere of dust, initially at rest, collapsing under the action of its own gravity; therefore the density $\rho$ is a function of time only and the pressure gradients are identically zero (this won't be valid anymore when later we introduce perturbations). Then we implement the modified algebra (4) to show how the singularity is removed and also derive some bounds on the deformation parameter $\mu$ by requiring that the non-relativistic assumption holds.

### 3.1 Hamiltonian formulation of the hunter model

Given spherical symmetry, it is enough to study the evolution of the radius $r$ of the sphere; using the Newtonian gravitational potential, the Hamiltonian (actually the Hamiltonian per unit mass) results to be

$$\mathcal{H} = \frac{p^2}{2} - \frac{GM}{r}, \tag{6}$$

where $p$ is the momentum conjugate to $r$, $G$ is Newton's gravitational constant, and $M$ is the total mass of the cloud. The Hamilton equations are

$$\dot{r} = \frac{\partial \mathcal{H}}{\partial p} = p, \qquad \dot{p} = -\frac{\partial \mathcal{H}}{\partial r} = -\frac{GM}{r^2}; \tag{7}$$

dividing the second equation by the first we obtain a differential equation for $p(r)$ that can be integrated with the initial conditions $r = r_0$ and $\dot{r} = p = 0$ at $t = 0$, where $r_0$ is the initial radius of the cloud:

$$\frac{\partial p}{\partial r} = \frac{\dot{p}}{\dot{r}} = -\frac{GM}{r^2 p}, \quad p(r) = \pm\sqrt{2GM\left(\frac{1}{r} - \frac{1}{r_0}\right)}. \tag{8}$$

Then, substituting this in the equation for $\dot{r}$ with the minus sign (since in a collapse $\dot{r} < 0$) and defining $a = r/r_0$, we can obtain a solution for $a(t)$ in implicit form:

$$\dot{a} = -\sqrt{\frac{2GM}{r_0^3}\frac{1-a}{a}}, \tag{9}$$

$$\sqrt{\frac{2GM}{r_0^3}}\,t = \sqrt{a(1-a)} + \mathrm{acos}\sqrt{a}. \tag{10}$$

By setting $a(t_0) = 0$, we can find the time of collapse $t_0$ to be

$$t_0 = \frac{\pi}{2}\sqrt{\frac{r_0^3}{2GM}}. \tag{11}$$

The solution is shown in Fig. 1 compared with the modified non-singular solution that we will now derive.

### 3.2 Modified non-singular collapse

To obtain the modified evolution, we start from the same Hamiltonian (6) but derive the equations of motion through the modified Poisson brackets (4):

$$\dot{r} = p\left(1 - \frac{\mu^2 p^2}{c^2}\right), \qquad \dot{p} = -\frac{GM}{r^2}\left(1 - \frac{\mu^2 p^2}{c^2}\right), \tag{12}$$

where $p$ has the dimensions of a velocity and therefore we inserted the speed of light $c$ to keep $\mu$ dimensionless. Now, dividing the second equation by the first we obtain the same relation (8), and substituting we get a differential equation for $a(t)$ of the form

$$\dot{a} = -\sqrt{\frac{2GM}{r_0^3}\frac{1-a}{a}}\left(1 - \frac{2GM\mu^2}{r_0 c^2}\frac{1-a}{a}\right). \tag{13}$$

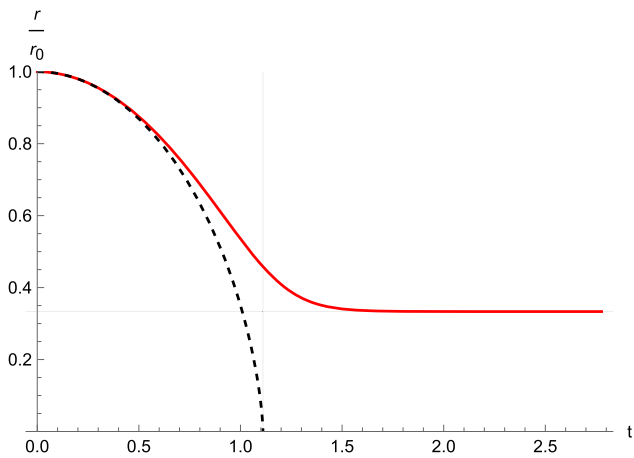

**Fig. 1** Comparison between the classical collapse (dashed black line) and the modified non-singular evolution (red continuous line) for generic values of the parameters; the collapse time $t_0$ and the minimum value $a_\infty$ are highlighted by faded grey lines

We see that the modified algebra has introduced a critical point: we find the value $a_\infty < 1$ such that $\dot{a} = 0$ as

$$1 - c_\mu \frac{1 - a_\infty}{a_\infty} = 0, \quad a_\infty = \frac{c_\mu}{1 + c_\mu}, \quad (14)$$

where we defined $c_\mu = 2GM\mu^2/r_0 c^2$ to shorten the notation. The solution for $a(t)$ can again be found only in implicit form:

$$\sqrt{\frac{2GM}{r_0^3}} t = \frac{\sqrt{a(1-a)}}{1 + c_\mu} + \frac{1 + 3c_\mu}{(1 + c_\mu)^2} \mathrm{acos}\sqrt{a}$$
$$+ \frac{2c_\mu^{\frac{3}{2}}}{(1 + c_\mu)^{\frac{5}{2}}} \sqrt{1 + 2b_-} \, (1 + b_+) \mathrm{atanh}\left(\frac{\sqrt{a} - 1}{\sqrt{(1-a)(1 + 2b_-)}}\right) +$$
$$- \frac{2c_\mu^{\frac{3}{2}}}{(1 + c_\mu)^{\frac{5}{2}}} \sqrt{1 + 2b_+} \, (1 + b_-) \mathrm{atanh}\left(\frac{\sqrt{a} - 1}{\sqrt{(1-a)(1 + 2b_+)}}\right),$$
$$(15)$$

where $b_\pm = c_\mu \pm \sqrt{c_\mu(1 + c_\mu)}$. It is trivial to see that in the limit $\mu \to 0$ we have $c_\mu, b_\pm \to 0$ and the standard solution (10) is recovered; it is also easy to verify that, when $a = a_\infty$, the arguments of both inverse hyperbolic tangents become 1 and the right-hand-side diverges, meaning that the inverse function $a(t)$ has an horizontal asymptote such that $a \to a_\infty$ when $t \to \infty$. In Fig. 1 the classical and the modified solutions are compared.

We highlight that this construction is similar to others where, through the implementation of alternative quantization procedures (or their corresponding semiclassical limits), a static solution was found that was not present with the standard framework [51,52].

At this point we can find some constraints on the deformation parameter $\mu$ by requiring that the Newtonian description be valid. In particular, we impose that the minimum radius be much greater than the Scharzschild radius $r_S = 2GM/c^2$: the condition $a_\infty \gg a_S$ implies

$$\frac{c_\mu}{1 + c_\mu} \gg a_S = \frac{r_S}{r_0}, \quad \mu \gg \sqrt{\frac{1}{1 - a_S}} \quad (16)$$

(note that $c_\mu = a_S \mu^2$). Therefore we find that for a cloud with initial mass and radius equal to those of our Sun we have

$$\mu \gg 1, \quad (17)$$

which was expected since for the Sun $a_S \sim 10^{-4}$ and the square root is basically 1; this relation may of course vary for different values of initial radius and mass, but even for more compact objects with $a_S \sim 2/5$ we would have $\mu \gg 1.3$.

As a secondary check, we require that the maximum speed reached during the collapse be non-relativistic. The maximum speed is found by setting $\ddot{r} = 0$ and substituting in $\dot{r}$: we find

$$\mu \gg \frac{2}{3\sqrt{3}} \sim 0.4, \quad (18)$$

which is slightly smaller but still of order 1. Note how this constraint, differently from the previous one, does not depend on any parameter. Therefore, we can conclude that, by taking the deformation parameter just one or two orders of magnitude greater than 1, we are assured that the Newtonian dynamics is still a good description for this model and that the collapse stops before the formation of a horizon. The obtained values are also compatible with previous attempts at finding values or limits on the GUP deformation parameter [53,54].

## 4 Non-relativistic perturbations

Let us now study the behaviour of perturbations in the Newtonian description. We will see that, for the non-singular case, a Jeans-like length naturally emerges. Note that, while the background configuration is determined by including cut-off physics effects, the evolution of the perturbations follows standard dynamics; this choice is justified by the observation that, while the background evolution is non-perturbatively sensitive to the cut-off physics, the smallness of the perturbations ensures that their dynamics can be satisfactorily described via standard gravity effects.

Still following Hunter [36], for the description of perturbations it is better to use an Eulerian representation. The system is then described by the following quantities:

$$\mathbf{v} = (r_0 \dot{a}, 0, 0), \quad (19a)$$
$$\rho = \rho_0 a^{-3}, \quad (19b)$$

$$\Phi = -2G\pi\rho r_0^2 \left(1 - \frac{a^2}{3}\right), \tag{19c}$$

where **v** is the velocity vector, $\rho$ and $\rho_0$ are the density of the cloud and its initial value, and $\Phi$ is the gravitational potential. These quantities are linked by the continuity, Euler and Poisson equations [40]:

$$\dot{\rho} + \nabla \cdot (\rho \mathbf{v}) = 0, \tag{20a}$$

$$\dot{\mathbf{v}} + \left(\dot{\mathbf{v}} \cdot \nabla\right)\mathbf{v} = -\nabla\Phi - \frac{\nabla P}{\rho}, \tag{20b}$$

$$\nabla^2 \Phi = 4\pi G \rho, \tag{20c}$$

where $P = P(\rho)$ is the pressure that depends only on the density due to the barotropic assumption. Now we perturb the quantities (19) to first order (higher-order corrections were investigated by Hunter later in [55,56]) as $\overline{\mathbf{v}} = \overline{\mathbf{v}} + \delta\overline{\mathbf{v}}$, $\rho = \overline{\rho} + \delta\rho$, $\Phi = \overline{\Phi} + \delta\Phi$, where the unperturbed quantities (those with the overline) already satisfy equations (20). By substituting into the Euler equation (20b) and taking the rotor, we obtain an equation for the vorticity $\delta \mathbf{w} = \nabla \times \delta v$:

$$\dot{\delta \mathbf{w}} = -\nabla \times (\delta \mathbf{w} \times \mathbf{v}), \tag{21}$$

with solution

$$\delta \mathbf{w} = \left(\frac{w_r}{a^2} + W, \frac{w_\theta}{a^2}, \frac{w_\varphi}{a^2}\right), \tag{22}$$

where $w_r, w_\theta, w_\phi$ and $W$ are arbitrary functions in spherical coordinates which must satisfy $\nabla \cdot \delta\mathbf{w} = 0$ (the divergence of a curl is identically zero in any system of coordinates); note that $W$ can be ignored since it represents a static distribution. Substituting this result back in equations (20), eliminating $\delta \Phi$ and using the polytropic relation $P = \kappa\rho^\gamma$, we obtain a term involving the Laplacian of $\delta\rho$ (for more details see [36,57]); in order to get rid of it, we can separate the variables as

$$\delta\rho(t, r_0 a, \theta, \varphi) = \delta\varrho(t)\,\psi(r_0 a, \theta, \varphi), \tag{23}$$

and then exploit the spherical symmetry of the problem by choosing $\psi$ to be an eigenfunction of the Laplacian operator:

$$\psi_{klm}(r_0 a, \theta, \varphi) = \Big(A_{lm}\,j_l(kr_0 a) \\ + B_{lm}\,y_l(kr_0 a)\Big) Y_{lm}(\theta, \varphi), \tag{24}$$

where $j_l$ and $y_l$ are spherical Bessel functions of the first and second kind, $A_{lm}$ and $B_{lm}$ are constant coefficients, and $Y_{lm}$ are spherical harmonics; this way we can write $\nabla^2 \delta\rho = -k^2 \delta\rho$, simplify the spatial part $\psi$ and obtain a differential equation just for the time dependent part of the density perturbation $\delta\varrho(t)$:

$$a^3 \dddot{\delta\varrho} + 8a^2 \dot{a} \ddot{\delta\varrho} \\ + \left(-4\pi\rho_0 G + k^2 v_s^2 a + 12 a \dot{a}^2 + 3 a^2 \ddot{a}\right) \delta\varrho = 0, \tag{25}$$

where $v_s^2 = \partial P/\partial \rho = \kappa\gamma\rho^{\gamma-1}$.

Until now, no reference to any solution was made. At this point we can insert the different expressions for $a(t)$ and its derivatives to obtain the solution $\delta\varrho(t)$ for the two cases.

### 4.1 The singular classical case

In the standard case we have

$$\dot{a} = -\sqrt{\frac{2GM}{r_0^3}\frac{1-a}{a}}, \quad \ddot{a} = -\frac{GM}{r_0^3 a^2}, \tag{26}$$

so the perturbation equation (25) becomes

$$a^3 \dddot{\delta\varrho} + 8a^2 \dot{a} \ddot{\delta\varrho} \\ + \left(\frac{6GM}{r_0^3}(3 - 4a) + k^2 v_0^2 a^{4-3\gamma}\right)\delta\varrho = 0, \tag{27}$$

where $v_0^2 = \kappa\gamma\rho_0^{\gamma-1}$. Now, since we are interested in the asymptotic behaviour near the singularity, we can take the solution (10) and perform an asymptotic expansion for $t \to t_0$, $a \to 0$, where $t_0$ is the collpase time given by Eq. (11), obtaining the explicit expression

$$a^{\text{asymp}}(t) = \left(\frac{3}{4}\pi\right)^{\frac{2}{3}} \left(1 - \frac{t}{t_0}\right)^{\frac{2}{3}}. \tag{28}$$

After some manipulation, substituting this in Eq. (27) yields

$$y^2 \frac{d^2 \delta\varrho}{d^2 y} - \frac{16}{3} y \frac{d\delta\varrho}{dy} \\ + \left(4 + \frac{3^{\frac{2}{3}-2\gamma} \pi^{\frac{8}{3}-2\gamma}}{2^{\frac{13}{3}-4\gamma}}\frac{k^2 v_0^2 r_0^3}{GM} y^{\frac{8}{3}-2\gamma}\right)\delta\varrho = 0, \tag{29}$$

where $y = 1 - t/t_0$ so that the limit $t \to t_0$ corresponds to $y \to 0$; the general solution is

$$\delta\varrho(y) = A_+ f_+(y) + A_- f_-(y), \tag{30}$$

$$f_\pm(y) = \frac{J_{\frac{\pm 5}{8-6\gamma}}\left((\alpha y)^{\frac{4}{3}-\gamma}\right)}{y^{\frac{13}{6}}}, \tag{31}$$

where $A_\pm$ are integration constants, $\alpha$ is a dimensionless constant containing the parameters $v_0, k, r_0$ and $M$, and $J_n$ is the Bessel function of the first kind. Studying the asymptotic behaviour, for $1 \leq \gamma < 4/3$ we have

$$f_+ \sim y^{-3}, \quad f_- \sim y^{-\frac{4}{3}}, \tag{32}$$

while for $4/3 < \gamma \leq 5/3$ we have

$$f_\pm \sim \frac{\cos((\alpha y)^{\frac{4}{3}-\gamma})}{y^{\frac{17}{6}-\frac{\gamma}{2}}}; \tag{33}$$

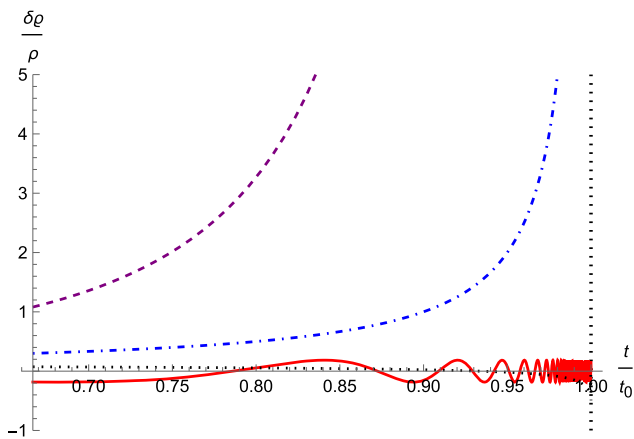

**Fig. 2** The asymptotic behaviour of the density contrast $\delta\varrho/\overline{\rho}$ for different values of the polytropic index in the standard singular case: $1 \leq \gamma < \frac{4}{3}$ (blue dot-dashed line), $\gamma = \frac{4}{3}$ (purple dashed line), $\frac{4}{3} < \gamma < \frac{5}{3}$ (black dotted line), $\gamma = \frac{5}{3}$ (red continuous line). The parameters have been chosen to have as initial value $\delta\varrho/\overline{\rho} \approx 10^{-1}$

remembering that in the asymptotic regime we have $\overline{\rho} \propto a^{-3} \propto y^{-2}$, the density contrast $\delta\varrho/\overline{\rho}$ will behave in the following ways:

$$\frac{\delta\varrho}{\overline{\rho}} \propto \frac{1}{y} \quad \text{for} \quad 1 \leq \gamma < \frac{4}{3}, \tag{34a}$$

$$\frac{\delta\varrho}{\overline{\rho}} \propto \frac{1}{y^{\frac{13}{6}}} \quad \text{for} \quad \gamma = \frac{4}{3}, \tag{34b}$$

$$\frac{\delta\varrho}{\overline{\rho}} \propto \frac{\cos\left((\alpha y)^{\frac{4}{3}-\gamma}\right)}{y^{\frac{5}{6}-\frac{\gamma}{2}}} \quad \text{for} \quad \frac{4}{3} < \gamma < \frac{5}{3}, \tag{34c}$$

$$\frac{\delta\varrho}{\overline{\rho}} \propto \cos\left(\frac{1}{(\alpha y)^{\frac{1}{3}}}\right) \quad \text{for} \quad \gamma = \frac{5}{3}. \tag{34d}$$

Therefore we conclude that, except for the last case where the frequency of oscillations increases but the amplitude remains constant, the perturbations collapse faster than the background and a fragmentation process is favoured [58]. The behaviour of $\delta\varrho/\overline{\rho}$ for different values of $\gamma$ is depicted in Fig. 2.

### 4.2 The modified non-singular case

In the modified model, the expressions for the derivatives of $a$ are different:

$$\dot{a} = -\sqrt{\frac{2GM}{r_0^3}\frac{1-a}{a}}\left(1 - c_\mu \frac{1-a}{a}\right), \tag{35a}$$

$$\ddot{a} = -\frac{GM}{r_0^3 a^2}\left(1 - c_\mu \frac{1-a}{a}\right)\left(1 - 3c_\mu \frac{1-a}{a}\right), \tag{35b}$$

so the differential equation for the amplitude $\delta\varrho$ of the perturbations is more complicated; however, we are aided by the fact that the asymptotic behaviour at lowest order is just

$a(t) \to a_\infty, \dot{a}, \ddot{a} \to 0$. Therefore the perturbation equation (25) becomes simply

$$a_\infty^3 \ddot{\delta\varrho} + \left(k^2 v_0^2 a_\infty^{4-3\gamma} - \frac{3c^2 a_\infty}{2(1-a_\infty)r_0^2 \mu^2}\right)\delta\varrho = 0, \tag{36}$$

and the solution is a simple sum of two exponential functions:

$$\delta\varrho(t) = B_+ e^{+\lambda t} + B_- e^{-\lambda t}, \tag{37}$$

$$\lambda = \sqrt{\frac{3c^2}{2(1-a_\infty)a_\infty^2 r_0^2 \mu^2} - k^2 v_0^2 a_\infty^{1-3\gamma}}; \tag{38}$$

therefore the behaviour of perturbations depends entirely on the sign of the quantity in the square root. First of all, we can compute the value of $v_0$ (corresponding to the speed of sound at the start of the collapse) using a quasi-static approximation:

$$-\frac{GM^2}{r_0^2} + 4\pi r_0^2 P_0 = 0, \tag{39}$$

where $P_0$ is the pressure at the start of the collapse, thus obtaining

$$v_0^2 = \kappa\gamma\rho_0^{\gamma-1} = \gamma\frac{P_0}{\rho_0} = \frac{\gamma}{3}\frac{GM}{r_0}; \tag{40}$$

then, from the expression of $a_\infty$ we can rewrite the value of $\lambda$ as

$$\lambda = \sqrt{\frac{3GM}{r_0^3 a_\infty^3}\left(1 - \frac{\gamma}{9}k^2 r_0^2 a_\infty^{4-3\gamma}\right)}. \tag{41}$$

Now, when $\lambda = 0$, we obtain a pivot scale $k_0$ of the form

$$k_0 = \frac{3}{r_0}\sqrt{\frac{a_\infty^{3\gamma-4}}{\gamma}} \tag{42}$$

such that we can rewrite $\lambda$ as

$$\lambda = \sqrt{\frac{3GM}{r_0^3 a_\infty^3}\left(1 - \frac{k^2}{k_0^2}\right)}. \tag{43}$$

Therefore, for $k < k_0$, we have $\lambda^2 > 0$ so $\delta\varrho$ and $\delta\varrho/\overline{\rho}$ diverge while, for $k > k_0$, $\lambda^2 < 0$ so $\delta\varrho$ oscillates with constant amplitude and the density contrast $\delta\varrho/\overline{\rho}$ is ultimately damped to zero. This translates to a Jeans-like length scale $\ell_0 = 2\pi/k_0$ of the form

$$\ell_0 = \frac{2\pi r_0}{3}\sqrt{\frac{\gamma}{a_\infty^{3\gamma-4}}}, \tag{44}$$

above which a perturbation diverges and the fragmentation process is initiated while below it the perturbation is damped and erased. Note that, since $0 < a_\infty < 1$ and it is constant, for each value of the polytropic parameter $\gamma$ we can find both behaviours depending only on the initial scale of the perturbation. However, for some values of $\gamma$ it turns out that the scale $\ell_0$ is bigger than the initial radius of the cloud, so all perturbations will disappear. Figure 3 shows the length scale $\ell_0$ as function of the deformation parameter $\mu$ for different values of $\gamma$ (the other parameters are again those of our Sun): we see that in order to have $\ell_0 < r_0$ and allow the fragmentation process, we must first of all have $\gamma < 4/3$ since, for $\gamma = 4/3$, $\ell_0$ does not depend on $a_\infty$ (and therefore on $\mu$) and is a constant already greater than $r_0$; this upper limit is further reduced by the condition $r_\infty \gg r_S$, and therefore we must have $1 \leq \gamma < \gamma_1 < 4/3$, where $\gamma_1$ is such that $\ell_0 = r_0$ at a the value of $\mu$ for which $r_\infty = r_S$.

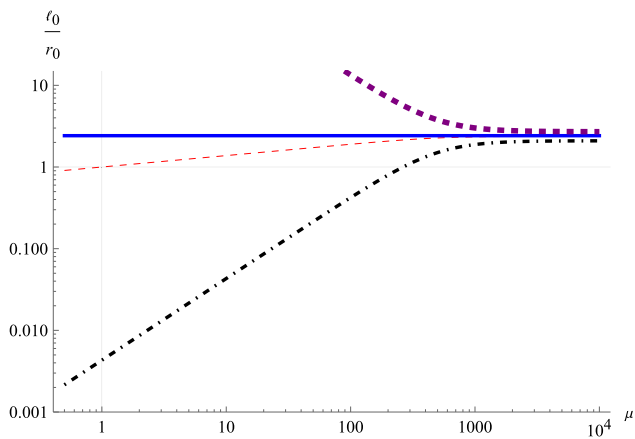

**Fig. 3** The Jeans-like length scale $\ell_0$ as function of the deformation parameter $\mu$ for different values of $\gamma$ in the modified non-singular model. From top to bottom: $\gamma = \frac{5}{3}$ (thick purple dashed line), $\gamma = \frac{4}{3}$ (continuous blue line) for which $\ell_0$ is constant, $\gamma = \gamma_1$ (thin red dashed line) that crosses the point (1, 1), and $\gamma = 1$ (black dot-dashed line); the faded gray lines correspond to $\ell_0 = r_0$ and $\mu = 1$

## 5 Relativistic gravitational collapse

In this section we study the collapse from a general relativistic point of view. Therefore the starting point will be the Oppenheimer–Snyder collapse model [37,59], for which we will present the Hamiltonian formulation and then implement on it the modified algebra (4).

### 5.1 The Oppenheimer–Snyder model and its Hamiltonian formulation

The Oppenheimer–Snyder (OS) model is the simplest and most widely known model of gravitational collapse. Its importance lies in highlighting the need to consider two different observers, one stationary outside the collapsing matter and one comoving with it. The original paper [37] starts from the outside Schwarzschild metric in the standard form

$$ds^2 = \left(1 - \frac{r_S}{R}\right)c^2 dT^2 - \frac{dR^2}{1 - \frac{r_S}{R}} - R^2 d\theta^2 - R^2 \sin^2(\theta) d\varphi^2, \tag{45}$$

where $T = T(t, r)$ and $R = R(t, r)$ are the external variables as functions of the internal $t$, $r$; then, by requiring spherical symmetry and homogeneity and implementing matching conditions on the surface $r_0$, it finds the equations that the internal metric must satisfy and computes the collapse time as seen from a comoving observer. The internal metric results to be that of a closed FLRW model [39]:

$$ds^2 = c^2 dt^2 - a^2(t)\left(\frac{dr^2}{1 - Kr^2} + r^2 d\theta^2 + r^2 \sin^2(\theta) d\varphi^2\right), \tag{46}$$

where $a(t)$ is the scale factor and $K > 0$ is the positive spatial curvature; while in the actual FLRW model the latter can always be set to $\pm 1$ by rescaling the variables, here it can be linked to the initial parameters of the cloud both through physical arguments [60] and through a comparison of the solutions, as we will see shortly.

To obtain the Hamiltonian formulation for the OS model one starts with the ADM-reduced action $S$ for spherically symmetric spacetimes [61–63] filled with Brown-Kuchař dust [64]; then, by implementing matching conditions between the Schwarzschild (45) and the FLRW (46) metrics and performing a partial symmetry reduction, the Hamiltonian gets split in three different contributions:

$$S = \int dt\, (p_a \dot{a} + P_\tau \dot{\tau} - N\mathcal{H} - M_+ \dot{T}_+) + \int dt \int_{r_0}^\infty dr (P_R \dot{R} + P_L \dot{L} - N^0 \mathcal{H}_0 - N^r \mathcal{H}_r), \tag{47}$$

$$\mathcal{H}_0 = \frac{P_L^2 L}{2R^2} - \frac{P_R P_L}{R} + \frac{R'^2 + 2RR''}{2L} - \frac{L'RR'}{L^2} - \frac{L}{2}, \tag{48a}$$

$$\mathcal{H}_r = P_R R' - P_L' L, \tag{48b}$$

$$\mathcal{H} = -\frac{\chi}{6V_S}\frac{p_a^2}{a} - \frac{3V_S}{2\chi}Kc^2 a + P_\tau, \tag{48c}$$

where $\tau$ is dust proper time, $a$ is the scale factor for the internal metric, $L$ and $R$ are the functions appearing in the spherically symmetric external metric (and can be found by comparison with (45)), $N$, $N^0$ and $N^r$ are Lagrange multipliers, the $P_i$ are the momenta conjugate to their respective variables, $M_+ \dot{T}_+$ is a boundary term containing the ADM mass and the Schwarzschild-Killing time at asymptotic infinity, $\mathcal{H}_0$ and $\mathcal{H}_r$ are the super-Hamiltonian and super-momentum for the exterior of the dust cloud, and $\mathcal{H}$ is the Hamiltonian for the interior.

We are interested of course in the internal Hamiltonian (48c): it contains the scale factor $a$, its conjugate momentum $p_a$, the spatial curvature $K$, the momentum conjugate to dust proper time $P_\tau$ which contains the energy density of the cloud, Einstein's constant $\chi = 8\pi G/c^2$ and the internal volume $V_S$ of the sphere given by

$$V_S = \int_0^{r_0} dr \frac{4\pi r^2}{\sqrt{1-Kr^2}}$$
$$= \frac{4\pi}{2K}\left(\frac{\mathrm{asin}(r_0\sqrt{K})}{\sqrt{K}} - r_0\sqrt{1-Kr_0^2}\right). \quad (49)$$

Had we started directly from the FLRW model we would have obtained a very similar Hamiltonian, with a (constant in the case of pressureless dust) energy density term instead of the (still constant) $P_\tau$.

The matching conditions imply the following identifications:

$$L = \frac{a}{\sqrt{1-Kr_0^2}}, \qquad R = ar_0; \quad (50)$$

then, by studying the dynamics of $a$ in the interval $0 < a \le 1$, the FLRW metric describes the interior of the dust cloud from the point of view of a comoving observer. After deriving the solution it will be sufficient to multiply the comoving scale factor $a$ by the initial radius $r_0$ to obtain the dynamics of the physical radius of the cloud; this is similar the usual FLRW description where, even though the scale factor is defined only up to a constant, its value today is taken to be $a_0 = 1$ in order to be able to find physical distances. In our description we will keep the scale factor as physical in order to keep the notation more compact, and only reintroduce the initial radius $r_0$ if needed for numerical purposes. For more information on the derivation of the action and the Hamiltonians see [65–68].

Focusing on the interior Hamiltonian, the equations of motion are

$$\dot{a} = -\frac{\chi}{3V_S}\frac{p_a}{a}, \quad (51a)$$
$$\dot{p}_a = -\frac{\chi}{6V_S}\frac{p_a^2}{a^2} + \frac{3Kc^2V_S}{2\chi}; \quad (51b)$$

similarly to the non-relativistic case, it is useful to compute $\dot{p}_a/\dot{a}$ in order to have an easily solvable differential equation for $p_a(a)$:

$$\frac{\partial p_a}{\partial a} = \frac{p_a}{2a} - \frac{9}{2}\frac{Kc^2V_S^2}{\chi^2}\frac{a}{p_a}, \quad (52)$$
$$p_a(a) = \frac{3cV_S}{\chi}\sqrt{a(1-a)K}, \quad (53)$$

where we used the standard initial conditions $a = 1$ and $p_a = 0$ at $t = 0$. Substituting this in Eq. (51a), we obtain the same differential equation (9) of the standard case, but with a different constant:

$$\dot{a} = -\sqrt{Kc^2\frac{1-a}{a}}; \quad (54)$$

therefore we already know the solution and furthermore we can identify the curvature as function of the initial parameters

of the cloud:

$$\sqrt{a(1-a)} + \mathrm{acos}\sqrt{a} = \sqrt{K}\,ct, \quad (55)$$
$$K = \frac{2GM}{c^2 r_0^3} = \frac{r_S}{r_0^3}; \quad (56)$$

this is the same identification found from physical arguments in [60], where the authors find a link between the Schwarschild and the FLRW metrics. With this identification we can also rewrite the expression of $V_S$ as

$$V_S = \frac{2\pi r_0^3}{\sqrt{a_S^3}}\left(\mathrm{asin}\sqrt{a_S} - \sqrt{a_S(1-a_S)}\right), \quad (57)$$

where we have again defined $a_S = r_S/r_0$.

The solution is shown later in Fig. 4, compared with the modified solution which we will now derive.

### 5.2 Non-singular relativistic collapse

To find the modified dynamics, we again start from the same Hamiltonian (48c) but use the modified algebra (4). The new equations of motion then are

$$\dot{a} = -\frac{\chi}{3V_S}\frac{p_a}{a}\left(1 - \frac{\mu^2 p_a^2}{\hbar^2}\right), \quad (58a)$$
$$\dot{p}_a = \left(-\frac{\chi}{6V_S}\frac{p_a^2}{a^2} + \frac{3Kc^2V_S}{2\chi}\right)\left(1 - \frac{\mu^2 p_a^2}{\hbar^2}\right), \quad (58b)$$

where $p_a$ has the dimensions of an action so we introduced a Planck constant to still have $\mu$ dimensionless; dividing the second equation by the first we obtain the same relation (53), so the final differential equation for $a(t)$ becomes

$$\dot{a} = -\sqrt{Kc^2\frac{1-a}{a}}\left(1 - g_\mu(1-a)a\right), \quad (59)$$

where we defined $g_\mu = K(3\mu cV_S/\hbar\chi)^2$. Already from here we see that there is still a critical point, but its expression is different from the Newtonian case:

$$1 - g_\mu(1-a_\infty)a_\infty = 0, \quad a_\infty = \frac{1}{2} \pm \frac{1}{2}\sqrt{1 - \frac{4}{g_\mu}}. \quad (60)$$

First of all we see that, in order for $a_\infty$ to be real we must have $g_\mu \ge 4$, otherwise we will still have the collapse $a \to 0$, and this will imply a lower limit on the deformation parameter $\mu$ as we will see later; secondly, when that condition is satisfied, we will always have $a_\infty \ge 1/2$ (the solution with the minus sign will never be reached in this model, but only by starting below it with a positive derivative). Now, imposing the condition $a_\infty \gg a_S$, we obtain the following lower limits for $\mu$:

$$\mu \gg \frac{8\pi\hbar\sqrt{r_S r_0^3}}{3cMV_S} = \frac{2\hbar}{c\rho_0 V_S\sqrt{K}} \quad \text{for} \quad r_S < \frac{r_0}{2}, \quad (61a)$$

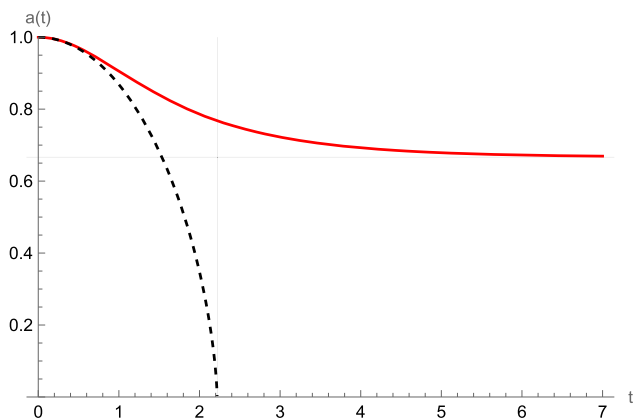

**Fig. 4** Comparison between the classical relativistic collapse (dashed black line) and the modified non-singular evolution (red continuous line) for generic values of the parameters; the collapse time $t_0$ and the minimum value $a_\infty$ are highlighted by faded grey lines; it is evident how $a_\infty > 1/2$

$$\mu \gg \frac{4\pi \hbar r_0^{\frac{5}{2}}}{3cMV_S\sqrt{r_0 - r_S}}$$

$$= \frac{\hbar\sqrt{K}}{c\rho_0 V_S \sqrt{a_S(1-a_S)}} \quad \text{for} \quad r_S > \frac{r_0}{2}; \tag{61b}$$

note that when $r_S < r_0/2$, we already have $a_\infty > a_S$ by construction; indeed, the constraint (61a) actually corresponds to the reality condition $g_\mu > 4$. When we insert the parameters of our Sun the two conditions yield the following lower limits for the parameter $\mu$:

$$\mu \gg 10^{-84} \quad \text{for } r_S < \frac{r_0}{2}, \tag{62a}$$

$$\mu \gg 2.5 \times 10^{-82} \quad \text{for } r_S > \frac{r_0}{2}. \tag{62b}$$

These small values are due to $g_\mu$ being a very large number. We can safely assume that the asymptotic radius of the cloud is always greater than its Schwarzschild radius as long as $\mu \neq 0$.

Now, the non-singular solution can again be expressed only in implicit form:

$$\frac{\sqrt{8}\,\mathrm{atan}\left(\sqrt{\frac{2}{d_-}\frac{1-a}{a}}\right)}{\sqrt{g_\mu(g_\mu - 4)d_-}} - \frac{\sqrt{8}\,\mathrm{atan}\left(\sqrt{\frac{2}{d_+}\frac{1-a}{a}}\right)}{\sqrt{g_\mu(g_\mu - 4)d_+}} = \sqrt{K}\,ct, \tag{63}$$

where we defined $d_\pm = 2 - g_\mu \pm \sqrt{g_\mu(g_\mu - 4)}$ to shorten the notation. The solution is presented in Fig. 4, compared with the unmodified relativistic evolution (55).

## 6 Relativistic perturbations

We will now study the behaviour of density perturbations in the relativistic setting. We will mainly follow [41], meaning that we will study linear perturbations of the Einstein equations.

First of all it is convenient to rewrite the FLRW metric in an easier form, introducing conformal time $\eta$ and the new variable $X$ defined as

$$d\eta = \frac{dt}{a}, \tag{64}$$

$$dX^2 = \frac{dr^2}{1 - Kr^2}, \quad X = \frac{\mathrm{asin}(\sqrt{K}\,r)}{\sqrt{K}}, \tag{65}$$

where the expression of $X$ as function of $r$ is valid for $K > 0$; this way, the FLRW metric inside of the cloud rewrites as

$$ds^2 = a^2(\eta)\Big(c^2 d\eta^2 - dX^2 - r^2 d\theta^2 - r^2 \sin^2(\theta)\, d\varphi^2\Big), \tag{66}$$

$$r = r(X) = \frac{\sin(\sqrt{K}\,X)}{\sqrt{K}}. \tag{67}$$

Small perturbations are described by changes in the metric tensor, in the four-velocity and in the scalar density, parametrized as $g_{jk} = \bar{g}_{jk} + \delta g_{jk}$, $u^j = \bar{u}^j + \delta u^j$ and $\rho = \bar{\rho} + \delta \rho$. Without loss of generality we can impose the synchronous gauge, thus setting $\delta g_{00} = \delta g_{0\alpha} = 0$ (latin indices go from 0 to 3, while greek indices refer to the spatial part and therefore go from 1 to 3). If the unperturbed system is comoving, we can set $u^\alpha = 0$ and $u^0 = 1/a$, and then from the unitarity of the four-velocity we obtain $\delta u^0 = 0$. Perturbations of the metric tensor imply perturbations of the Ricci tensor $R_j^k$ and of the Ricci scalar $R$ of the form

$$\delta R_\alpha^\beta = \frac{1}{2a^2}\left(\delta g_\alpha^{\gamma;\beta}{}_{;\gamma} + \delta g^\beta{}_{\gamma;\alpha}{}^{;\gamma} - \delta g_\alpha^{\beta;\gamma}{}_{;\gamma} - \delta g^{;\beta}_{;\alpha}\right)$$
$$+ \frac{1}{c^2a^2} \tag{68a}$$
$$\times \left(\frac{1}{2}\delta g_\alpha^{\beta\,\prime\prime} + \frac{a'}{a}\delta g_\alpha^{\beta\,\prime} + \frac{a'}{2a}\delta g' \delta_\alpha^\beta - 2Kc^2 \delta g_\alpha^\beta\right),$$

$$\delta R_0^0 = \frac{1}{2a^2}\left(\delta g'' + \frac{a'}{a}\delta g'\right), \tag{68b}$$

$$\delta R_\alpha^0 = \frac{1}{2a^2}\left(\delta g'_{;\alpha} - \delta g^\beta{}_\alpha{}'_{;\beta}\right), \tag{68c}$$

$$\delta R = \frac{1}{a^2}\left(\delta g_\alpha^{\gamma;\alpha}{}_{;\gamma} - \delta g^{;\alpha}_{;\alpha}\right)$$
$$+ \frac{1}{c^2a^2}\left(\delta g'' + 3\frac{a'}{a}\delta g' - \frac{2Kc^2}{a}\delta g\right), \tag{68d}$$

where $\delta g$ is the trace of the metric perturbations, $\delta_\alpha^\beta$ is Kronecker's delta, a semicolon indicates a covariant derivative and a prime a derivative after $\eta$. On the other hand, we can write the perturbed components of the Energy–Momentum

tensor $T_j{}^k$ as

$$\delta T_j{}^k = (P + c^2\rho)(u_j \delta u^k + u^k \delta u_j)$$
$$+ (\delta P + c^2 \delta\rho) u_j u^k + \delta_j^k \delta P, \quad (69a)$$

$$\delta T_\alpha{}^\beta = -\delta_\alpha^\beta \frac{dP}{d\rho} \frac{\delta T_0{}^0}{c^2}, \quad (69b)$$

$$\delta T_0{}^\alpha = -a(P + \rho c^2) \delta u^\alpha, \quad (69c)$$

$$\delta T_0{}^0 = -c^2 \delta\rho, \quad (69d)$$

where we have made use of the relation $\delta P = (dP/d\rho)\delta\rho$. In the linear approximation, the perturbations satisfy the equation

$$\delta R_j{}^k - \frac{1}{2} \delta_j^k \delta R = \frac{\chi}{c^2} \delta T_j{}^k, \quad (70)$$

which yields the following equations for the perturbations of the metric:

$$\left(\delta g_\alpha{}^{\gamma;\beta}{}_{;\gamma} + \delta g_{\gamma;\alpha}^{\beta\ ;\gamma} - \delta g_\alpha^{\beta;\gamma}{}_{;\gamma} - \delta g_{;\alpha}^{;\beta}\right)$$
$$+ \frac{1}{c^2}\left(\delta g_\alpha^{\beta\ \prime\prime} + 2\frac{a'}{a}\delta g_\alpha^{\beta\ \prime} - Kc^2 \delta g_\alpha^\beta\right) = 0, \quad (71a)$$

$$\frac{1}{2}(\delta g_{;\gamma}^{;\gamma} - \delta g_\gamma^{\delta;\gamma}{}_{;\delta}) - \frac{1}{c^2}\left(\delta g'' + 2\frac{a'}{a}\delta g' - Kc^2 \delta g\right)$$
$$= \frac{3}{c^4} \frac{dP}{d\rho}\left(\frac{1}{2}(\delta g_\gamma^{\delta;\gamma}{}_{;\delta} - \delta g_{;\gamma}^{;\gamma}) + \frac{a'}{a}\delta g' - Kc^2 \delta g\right). \quad (71b)$$

Putting everything together, the final equation for the density perturbations turns out to be

$$\chi c^2 \delta\rho = \frac{1}{2a^2}\left(\delta g_\alpha^{\beta;\alpha}{}_{;\beta} - \delta g_{;\alpha}^{;\alpha} + \frac{2a'}{c^2 a}\delta g' - 2K \delta g\right). \quad (72)$$

For more details on the derivation of these expressions, see [41].

Now, any perturbation in a hyperspherical geometry such as the positively-curved FLRW model can be expanded in four-dimensional spherical harmonics (similarly to the expansion in three-dimensional spherical harmonics performed in the non relativistic case in Sect. 4). The scalar hyperspherical harmonics $Q^n$ can be expressed as [69];

$$Q^n = \sum_{l=0}^{n-1} \sum_{m=-l}^{l} A_{lm}^n Y_{lm}(\theta, \varphi) \Pi_{nl}(X), \quad (73)$$

$$\Pi_{nl} = \sin^l(\sqrt{K} X) \frac{d^{l+1} \cos(n\sqrt{K} X)}{d \cos(\sqrt{K} X)}^{l+1}, \quad (74)$$

where $l$ can only go from 0 to $n-1$, $A_{lm}^n$ are constant coefficients and $Y_{lm}$ are the standard three-dimensional spherical harmonics. As an example, the most symmetric hyperspherical harmonics with $l = 0$ take the form

$$Q^n = \frac{\sin(n\sqrt{K} X)}{\sin(\sqrt{K} X)}. \quad (75)$$

From here on we will drop the superscript $n$ to avoid cluttering the notation. All hyperspherical harmonics are scalar eigenfunctions of the Laplacian operator on the surface of a hypersphere with unit radius, and therefore they satisfy the relation

$$Q_{;\alpha}^{;\alpha} = -(n^2 - 1)Q. \quad (76)$$

Here the order $n$ of the harmonics is an integer, and will play a similar role to the wave number $k$ of the non-relativistic perturbations; it can be roughly interpreted as "the number of wavelengths that fit inside the radius of the sphere" i.e. as the ratio of the radius of the sphere to the length scale of a given perturbation. Note that there exist also vector and tensor hyperspherical harmonics, but they are not needed to study density perturbations.

Now, from the scalar harmonics $Q$ it is possible to construct the following tensors and vectors with the following symmetries:

$$Q_\alpha^\beta = \frac{\delta_\alpha^\beta}{3} Q, \quad Q_\alpha^\alpha = Q, \quad (77a)$$

$$Z_\alpha = \frac{Q_{;\alpha}}{n^2 - 1}, \quad Z_\alpha^{;\alpha} = -Q, \quad (77b)$$

$$Z_\alpha^\beta = \frac{Q_{;\alpha}^{;\beta}}{n^2 - 1} + Q_\alpha^\beta, \quad Z_\alpha^\alpha = 0. \quad (77c)$$

Then we can define

$$\delta g_\alpha^\beta = \Lambda(\eta) Z_\alpha^\beta + \Omega(\eta) Q_\alpha^\beta, \quad \delta g = \Omega Q, \quad (78)$$

so that the whole spatial evolution is contained within the two tensors $Q_\alpha^\beta$ and $Z_\alpha^\beta$ while the time evolution i.e. the amplitude is just given by the two functions $\Lambda$ and $\Omega$; now the equation for the density perturbations becomes

$$\chi c^2 \delta\rho = \frac{Q}{3a^2}\left(Kc^2(n^2 - 4)(\Lambda + \Omega) + 3\frac{a'}{a}\Omega'\right). \quad (79)$$

Inserting expression (78) into equations (71), we obtain two differential equations for the two functions $\Lambda$ and $\Omega$:

$$\Lambda'' + 2\frac{a'}{a}\Lambda' - \frac{Kc^2}{3}(n^2 - 1)(\Lambda + \Omega) = 0, \quad (80a)$$

$$\Omega'' + \left(2 + \frac{3}{c^2}\frac{dP}{d\rho}\right)\frac{a'}{a}\Omega'$$
$$+ \frac{Kc^2}{3}(n^2 - 4)(\Lambda + \Omega)\left(1 + \frac{3}{c^2}\frac{d}{dP}\rho\right) = 0. \quad (80b)$$

Note that in the relativistic context we cannot use the polytropic relation because it is not a solution of the relativistic continuity equation; in this case we will make an isothermal assumption and leave the speed of sound $v_s^2 = dP/d\rho$ as a free constant parameter.

It is important to consider that only harmonics with $n > 2$ correspond to physical perturbations. For $n = 1, 2$ the tensor $Z_\alpha^\beta$ cannot be constructed, and therefore it is necessary to put $\Lambda = 0$; then we are left with just a second-order equation for $\Omega$. When $n = 2$ both solutions for $\Omega$ can be ruled out by a transformation of the coordinates. When $n = 1$ only one of the two solutions can be ruled out by such a transformation; the second solution corresponds to a perturbation in the entire mass of the cloud, but space remains fully uniform and isotropic. Thus only $n > 2$ correspond to real physical perturbations of the metric. For more details, see [41].

Now we only need to insert the solutions for the scale factor $a$ in the two cases; however we first have to express them in terms of the new time variable $\eta$.

### 6.1 Classical Oppenheimer–Snyder Perturbations

In order to find the expression for $a(\eta)$, we go back to Eq. (54) and substitute $dt = a\, d\eta$, thus obtaining a differential equation in $\eta$ that is easily solved:

$$a' = -\sqrt{K\, c^2\, (1-a)\, a}, \quad a(\eta) = \cos^2\left(\frac{\sqrt{K}\, c\, \eta}{2}\right). \quad (81)$$

Now, Eq. (80) have two particular integrals that correspond to those fictitious change in the metric that can be ruled out by a transformation of the reference system; nevertheless, they are useful to lower the order of the two equations. The particular integrals are

$$\Lambda_1 = -\Omega_1 = \text{const.}, \quad (82a)$$

$$\Lambda_2 = -\sqrt{K}\, c\, (n^2 - 1) \int \frac{d\eta}{a}, \quad (82b)$$

$$\Omega_2 = \sqrt{K}\, c\, (n^2 - 1) \int \frac{d\eta}{a} - \frac{3a'}{\sqrt{K}\, c\, a^2}. \quad (82c)$$

At this point we can perform the following change of variables:

$$\Lambda + \Omega = (\Lambda_2 + \Omega_2)\sqrt{K}\, c \int \xi\, d\eta = -\frac{3a'}{a^2} \int \xi\, d\eta, \quad (83a)$$

$$\Lambda' - \Omega' = \sqrt{K}\, c\, (\Lambda_2' - \Omega_2') \int \xi\, d\eta + \sqrt{K}\, c\, \frac{\zeta}{a} =$$
$$= \left(3\left(\frac{a''}{a^2} - 2\frac{a'^2}{a^3}\right) - \frac{2Kc^2(n^2 - 1)}{a}\right) \quad (83b)$$
$$\int \xi\, d\eta + \sqrt{K}\, c\, \frac{\zeta}{a};$$

this way we obtain two coupled first-order differential equations for the new unknown functions $\xi$ and $\zeta$:

$$\xi' + \xi\left(\frac{2a''}{a'} + \frac{a'}{a}\left(\frac{3}{2c^2}\frac{dP}{d\rho} - 2\right)\right)$$
$$+ \frac{\sqrt{K}}{2c}\frac{dP}{d\rho}\zeta = 0, \quad (84a)$$

$$\zeta' + \left(1 + \frac{3}{2c^2}\frac{dP}{d\rho}\right)\frac{a'}{a}\zeta + \xi\left(-2\sqrt{K}\, c\, (n^2 - 1) + \frac{3}{\sqrt{K}\, c}\left(\frac{a''}{a} - \frac{2a'^2}{a^2} + \frac{3}{2c^2}\frac{a'^2}{a^2}\frac{dP}{d\rho}\right)\right) = 0. \quad (84b)$$

Now we have to perform asymptotic expansions; in particular, close to the singularity, the scale factor (81) behaves as

$$a^{\text{asymp}}(\eta) = \left(\frac{\pi}{2}\right)^2 \left(1 - \frac{\eta}{\eta_0}\right)^2, \quad \eta_0 = \frac{\pi}{\sqrt{K}\, c}, \quad (85)$$

where we have defined the time of singularity $\eta_0$. Then, inserting everything in Eq. (84) and introducing the velocity parameter $\beta = v_s/c < 1$, we obtain the following differential equations for $\xi$ and $\zeta$:

$$\frac{d\xi}{dx} + \frac{2 + 3\beta^2}{x}\xi - \sqrt{K}\, c\, \beta^2 \eta_0\, \zeta = 0, \quad (86a)$$

$$\frac{d\zeta}{dx} + \frac{2 + 3\beta^2}{x}\zeta + \frac{18(1 - \beta^2)}{\sqrt{K}\, c\, \eta_0^2 x^2}\xi = 0, \quad (86b)$$

where we defined $x = 1 - \eta/\eta_0$; the solutions are

$$\xi(x) = \frac{D_- x^{-\frac{\sigma}{2}} + D_+ x^{+\frac{\sigma}{2}}}{x^{\frac{5}{2} + 3\beta^2}}, \quad (87a)$$

$$\zeta(x) = \frac{\sqrt{K}\, c\, \eta_0 \left(D_-(\sigma + 1)x^{-\frac{\sigma}{2}} - D_+(\sigma - 1)x^{+\frac{\sigma}{2}}\right)}{36(1 - \beta^2)x^{\frac{3}{2} + 3\beta^2}}, \quad (87b)$$

$$\sigma = \sqrt{72\beta^4 - 72\beta^2 + 1}, \quad (87c)$$

where $D_\pm$ are integration constants. From these, we can obtain the expressions for $\Lambda$ and $\Omega$ and therefore for $\delta\rho$; remembering that $\bar{\rho} \propto a^{-3} \propto x^{-6}$, in the asymptotic limit $\eta \to \eta_0$ corresponding to $x \to 0$ we find the leading-term behaviour of the perturbations as

$$\frac{\delta\rho}{\bar{\rho}} \propto x^{-(\frac{5}{2} + 3\beta^2 + \frac{\sigma}{2})}. \quad (88)$$

Now, when $\sigma^2 > 0$ this quantity diverges; on the other hand, when $\sigma^2 < 0$ the exponent is complex, but it still has a (negative) real part so that, even if there are some oscillations, the amplitude is still divergent close to the singularity. For thoroughness, note that $\sigma^2 > 0$ for $\beta < \sqrt{(6 - \sqrt{34})/3}\,/2$ and $\beta > \sqrt{(6 + \sqrt{34})/3}\,/2$, which roughly correspond to 0.019 and 0.993 respectively (remember that by definition $0 < \beta < 1$).

To conclude, we have shown that in the relativistic case all perturbations can diverge and initiate the fragmentation process, differently from the Newtonian case where for $\gamma = 5/3$ the amplitude remained constant.

### 6.2 Non-singular relativistic perturbations

In the modified non-singular case, finding the expression for $a(\eta)$ is not necessary (although possible) because asymptotically the leading term is simply $a = a_\infty$ as in the Newtonian case. Therefore the density perturbations (79) will depend only on the sum $\Lambda + \Omega$, which can be found by summing Eq. (80) and solving them:

$$(\Lambda + \Omega)'' - K c^2 \left(1 - (n^2 - 4)\beta^2\right)(\Lambda + \Omega) = 0, \quad (89)$$

$$\Lambda(\eta) + \Omega(\eta) = E_+ e^{+\nu\sqrt{K}\,c\,\eta} + E_- e^{-\nu\sqrt{K}\,c\,\eta}, \quad (90)$$

$$\nu = \sqrt{1 - (n^2 - 4)\beta^2}, \quad (91)$$

where $E_\pm$ are constants of integration. Similarly to the Newtonian case, given that when $a' = 0$ the perturbations depend directly on the sum $\Lambda + \Omega$, the fate of the perturbations depends entirely on the nature of this parameter $\nu$ i.e. on the sign of the term inside the square root. Therefore, since only $n > 2$ are relevant for physical perturbations, for each value of $n$ there exist a critical value of $\beta$ such that above it all perturbations oscillate and are ultimately damped, while below it all perturbations diverge. Conversely, for each value of $\beta$, there exist a value of $n$ large enough such that above it all perturbations oscillate and are damped, while below it they all diverge. Given that a higher value of $n$ corresponds to a perturbation with a shorter scale, we have again found a Jeans-like length. If we want the collapse to be stable to all kinds of perturbations, we require $n$ to be the smallest possible i.e. $n = 3$, thus finding a lower limit on $\beta$:

$$\beta > \beta_0 = \frac{1}{\sqrt{5}} \approx 0.48. \quad (92)$$

## 7 Concluding remarks

We analyzed the gravitational collapse of a spherical dust configuration, both in the Newtonian limit and in the fully general relativistic case, by including in the dynamics cut-off physics effects. The particular modification we introduced consists of a generalized Heisenberg algebra inspired by Polymer Quantum Mechanics. We studied the collapse dynamics by replacing the standard Poisson brackets with those ones coming from the considered generalized approach, modulated, to some extent, from Loop Quantum Gravity [70–72]. Indeed there are some recent attempts to analyze the gravitational collapse of a dust cloud through effective Loop Quantum Gravity, and they mostly agree with the resolution of the singularity [73,74].

The very remarkable result we have drawn in both the two considered regimes has to be identified in the existence of a stable and asymptotically static configuration of the collapse, established at a radius greater than the Schwarzschild one. Furthermore, this feature takes place in correspondence to a sufficiently small value of the parameter accounting for the new cut-off physics. In other words, it is always possible to accommodate the stabilization of the gravitational collapse at super-Schwarzschild scales even when the deformation parameter is defined as a Planckian quantity, i.e. as regularizing physics only for very high energy scales.

We also obtained some specific constraints on the free equation of state parameters by requiring that the asymptotic configuration be stable, in particular for the relativistic isothermal case we arrived to the requirement that the sound velocity $\beta$ be greater than $1/\sqrt{5}$. This suggests that, even if the repulsive character of the modified gravitational dynamics creates a static macroscopic configuration also when matter pressure is negligible, the request that this configuration be also stable under small perturbations still requires that the elementary constituents of the collapsing gas have a significant free-streaming effect.

The present analysis must be regarded as the starting point for subsequent investigations in which the gravitational collapse is modelled in a realistic astrophysical context, in order to better understand the implications that the repulsive gravitational dynamics can have on the formation of compact objects. In particular, the impact of the repulsive effects on the equilibrium of a real relativistic star [6] is of interest in order to determine possible corrections to the mass limits in the proposed scenario.

**Acknowledgements** G. B. thanks the TAsP Iniziativa Specifica of INFN for their support.

**Data Availability Statement** This manuscript has no associated data or the data will not be deposited. [Authors' comment: The manuscript has no associated data.]

**Code Availability Statement** The manuscript has no associated code/software.